\documentclass[aps,prx,superscriptaddress,twocolumn,nofootinbib,10pt]{revtex4-2}
\usepackage[utf8]{inputenc}

\usepackage[titletoc,toc,title,page]{appendix}

\usepackage{csquotes}
\usepackage[italian,english]{babel}

\usepackage{microtype} 

\usepackage{bm} 

\usepackage{dsfont} 
\usepackage{amsmath,amssymb,amsthm,thmtools}
\usepackage{mathtools}
\usepackage{cases}
\usepackage{calc}
\usepackage{mathrsfs} 
\usepackage[normalem]{ulem} 

\usepackage{parskip}
\usepackage{nameref}
\usepackage[colorlinks=true]{hyperref}
\usepackage[nameinlink]{cleveref}
\crefname{appsec}{Appendix}{Appendices}
\crefname{box}{Box}{Box}
\hypersetup{
  colorlinks   = true, 
  urlcolor     = green!80!black, 
  linkcolor    = blue, 
  citecolor    = red!80!black 
}

\usepackage{physics} 

\usepackage{float} 

\usepackage{graphicx}

\usepackage[usenames,dvipsnames,table]{xcolor}
\usepackage{easyReview}

\usepackage{tikz}
\usetikzlibrary{calc,shapes.geometric}

\usepackage{placeins}
\usepackage{multirow,tabularx,booktabs}
\usepackage[most]{tcolorbox}
\newtcbtheorem{tbox}{Box}{enhanced, float*=t, width=\textwidth, label type=box}{box}

\usepackage[printonlyused,withpage,nohyperlinks,smaller]{acronym}

\graphicspath{{./figures/}}

\newcommand{\calD}{\mathcal{D}}
\newcommand{\calE}{\mathcal{E}}

\newcommand{\bs}[1]{\boldsymbol{#1}}

\begin{document}

\title{An operational definition of quantum information scrambling}

\author{Gabriele Lo Monaco}
\let\comma,
\affiliation{Universit\`a degli Studi di Palermo\comma{} Dipartimento di Fisica e Chimica - Emilio Segr\`e\comma{} via Archirafi 36\comma{} I-90123 Palermo\comma{} Italy}
\author{Luca Innocenti}
\let\comma,
\affiliation{Universit\`a degli Studi di Palermo\comma{} Dipartimento di Fisica e Chimica - Emilio Segr\`e\comma{} via Archirafi 36\comma{} I-90123 Palermo\comma{} Italy}
\author{Dario Cilluffo}
\affiliation{Institut f\"ur Theoretische Physik and IQST\comma{} Albert-Einstein-Allee 11\comma{} Universit\"at Ulm\comma{} 89069 Ulm\comma{} Germany}
\author{Diana~A.~Chisholm}
\let\comma,
\affiliation{Centre for Quantum Materials and Technologies\comma{} School of Mathematics and Physics\comma{} Queen’s University Belfast\comma{} BT7 1NN\comma{} United Kingdom}
\author{Salvatore Lorenzo}
\affiliation{Universit\`a degli Studi di Palermo\comma{} Dipartimento di Fisica e Chimica - Emilio Segr\`e\comma{} via Archirafi 36\comma{} I-90123 Palermo\comma{} Italy}
\author{G. Massimo Palma}
\affiliation{Universit\`a degli Studi di Palermo\comma{} Dipartimento di Fisica e Chimica - Emilio Segr\`e\comma{} via Archirafi 36\comma{} I-90123 Palermo\comma{} Italy}
\affiliation{NEST\comma{} Istituto Nanoscienze-CNR\comma{} Piazza S. Silvestro 12\comma{} 56127 Pisa\comma{} Italy}


\begin{abstract}
\noindent
Quantum information scrambling (QIS) is a characteristic feature of several quantum systems, ranging from black holes to quantum communication networks. While accurately quantifying QIS is crucial to understanding many such phenomena, common approaches based on the tripartite information have limitations due to the accessibility issues of quantum mutual information, and do not always properly take into consideration the dependence on the encoding input basis. To address these issues, we propose a novel and computationally efficient QIS quantifier, based on a formulation of QIS in terms of quantum state discrimination.
We show that the optimal guessing probability, which reflects the degree of QIS induced by an isometric quantum evolution, is directly connected to the \emph{accessible min-information}, a generalized channel capacity based on conditional min-entropy, which can be cast as a convex program and thus computed efficiently. By applying our proposal to a range of examples with increasing complexity, we illustrate its ability to capture the multifaceted nature of QIS in all its intricacy.
\end{abstract}

\maketitle

\section{Introduction}

The goal of numerous quantum secure communication protocols, like error-correcting codes, is to safeguard information from unauthorized access. This challenge is typically tackled by encoding the input information into highly entangled states of a system $S$: the encoding is designed in such a way an eavesdropper who can only access a portion of $S$ cannot retrieve the conveyed message. The input information is actually hidden into the correlations among different subsystem and only through a global measurement can the message be recovered. In other words, secure communication protocols {\it scramble} the information that needs to be protected.

Quantum information scrambling (QIS) is a characteristic feature of the dynamics of a plethora of quantum systems, ranging from spin chains to black holes~\cite{Hayden:2007cs,Sekino:2008he,maldacena2016bound}.
Despite the significant body of literature on the subject, devising reliable methods to witness and quantify QIS remains an active area of research. 
The two most common approaches to quantify QIS are out-of-time-ordered correlators (OTOCs)~\cite{larkin1969quasiclassical,swingle2016measuring,nakamura2019universal,bergamasco2019out,xu2022scrambling,garcia2021quantum,landsman2019verified,yoshida2019disentangling,ahmadi2022quantifying,hosur2016chaos,swingle2018unscrambling,lewis2019unifying,kidd2021saddle} on one hand, and entropic quantifiers~\cite{zhuang2022phase,alba2019quantum,touil2020quantum,touil2021information,yeung1991new,cerf1998information,kitaev2006topological,casini2009remarks,hayden2013holographic,rota2016tripartite,hosur2016chaos,schnaack2019tripartite,iyoda2018scrambling,pappalardi2018scrambling,seshadri2018tripartite} on the other.
We will focus in this paper on the latter approach, which lends itself to a more direct informational interpretation.
In particular, the tripartite information has been proposed and is often used as a witness of QIS~\cite{hosur2016chaos}.
In a previous work, we highlighted a number of limitations associated to this way of directly using the tripartite information to quantify QIS~\cite{lomonaco2023quantum} that make it a misleading quantifier in several situations, and introduced an alternative operational quantifier that, being based on directly experimentally accessible quantities, solves many of these issues.
In particular, this quantifier, named ``accessible tripartite information'', involves the calculation of accessible mutual information between input states and different locally measured output states, and thus involves an optimization over both input and output measurement bases.

In this paper, we make further strides towards the goal of devising a QIS quantifier that has a clear information-theoretic interpretation and remains easily computable in practice.
We focus in particular on the best approach to quantify correlations between input and locally measured output states.
To achieve this, we start with the core observation that QIS can be contextualized within the framework of quantum state discrimination~\cite{watrous2018theory,barnett2009quantum,bae2015quantum}, which is best suited to accurately quantify the degree of information locally retrievable from individual measurement outcomes after scrambling dynamics.
More specifically, the optimal probability to discriminate between an ensemble of states quantifies the information retrievable in the \textit{single-shot regime}, where information is encoded into single copies of the states sent through the channel.
This is in contrast with mutual-information-based quantities, which instead quantify how much information can be transmitted allowing to encode information into asymptotically many input states sent through multiple copies of the scrambling dynamics~\cite{cover1999elements}.
In turn, the optimal probability to discriminate the states in a given ensemble can be characterized in terms of \textit{single-shot information quantifiers} based on the min-entropy~\cite{konig2009operational,skrzypczyk2019robustness}.
In this language, the QIS of isometric dynamics corresponds to the observer's probability to correctly guess, from local measurements, which state was sent through the channel among a finite ensemble of possible ones.
For example, when the discrimination probability approaches one for some choice of local POVM, the information about the choice of input states can be effectively retrieved from local measurements, and thus no scrambling occurs.
Following this route, we show that the \textit{accessible min-information}~\cite{skrzypczyk2019robustness,takagi2019general} is well-suited to quantify the probability of discriminating input states from local measurements after isometric dynamics, and is therefore a faithful quantifier of QIS in the single-shot regime that does not require to perform a maximization with respect to input encodings.

{\it Outline ---} In section \ref{sec:issues_I3} we briefly review the most common quantifiers of QIS. In section \ref{sec:our_definition_of_QIS} we explore the relationship between the rates of information transmission through a quantum channel and quantum state discrimination. This leads us to propose a new approach to quantify QIS in terms of probability of discriminating among states of an ensembles. We show how this probability can be translated into a generalized channel capacity that can be efficiently computed by means of semi-definite convex programming techniques.
In Sections \ref{sec:QECC} and \ref{sec:Chains} we apply our proposal to two examples.


\section{Quantum Information Scrambling: an overview}
\label{sec:issues_I3}
To understand what quantum information scrambling is, it can be useful to consider the following general situation: some input information is encoded in the state $\rho_A\in \mathcal{D}_A$ of a system $A$, where $\mathcal{D}_A$ is the space of density matrices on $A$. The system $A$ interacts with the environment $B$ prepared in $\rho_B\in \mathcal{D}_B$ and the evolution is unitary. After the interaction, an observer tries to decode the input information by performing measurements on a subsystem $C\subset AB$ of size $k\equiv{\rm dim}\,\mathcal{H}_{C}$, where $\mathcal H_C$ denotes the Hilbert space of the subsystem $C$. If no subsystem of size $k$ allows to retrieve the input information, we say that $k$-scrambling occurred.
Note that specifying the dimension of the observed subsystem is crucial, as measuring sufficiently large subsystems $C$ one can always fully retrieve the information in $A$, being the evolution unitary.

A standard quantifier of QIS are OTOCs~\cite{larkin1969quasiclassical,swingle2016measuring,nakamura2019universal,bergamasco2019out,xu2022scrambling,garcia2021quantum,landsman2019verified,yoshida2019disentangling,ahmadi2022quantifying,hosur2016chaos,swingle2018unscrambling,lewis2019unifying,kidd2021saddle}, which are defined as:
\begin{equation}
{\rm OTOC}(t)=\mathbb{E}_{V,W}\langle[V,W(t)]^\dagger[V,W(t)]\rangle_{\beta}\,.
\end{equation}
Here, $V,W$ are operators acting locally on disjoint subsystems of $AB$; $\mathbb{E}_{V,W}$ is the average computed extracting $V,W$ from some ensemble; the expectation value $\langle \cdot\rangle_\beta$ is taken with respect to the thermal state $\rho_{\beta}\in \mathcal{D}_{AB}$ at inverse temperature $\beta$; and $W(t)=U^\dagger W U$ is the operator $W$ evolved in the Heisenberg picture.
A drawback of OTOCs is that they lack a clear information-theoretic interpretation.
More specifically, they do not provide any insight into where the input information is stored, how much information can be retrieved from a given subsystem. 
From this perspective, entropic-based quantifiers, such as those based on the tripartite information (TI)~\cite{zhuang2022phase,alba2019quantum,yeung1991new,cerf1998information,kitaev2006topological,casini2009remarks,hayden2013holographic,rota2016tripartite,hosur2016chaos,schnaack2019tripartite,iyoda2018scrambling,pappalardi2018scrambling,seshadri2018tripartite}, have a more direct interpretation and are thus preferable.
The TI is defined in terms of quantum mutual informations as
\begin{equation}
\label{eq:TI}
    I_3(R:C:D)=I(R:C)+I(R:D)-I(R:CD),
\end{equation}
where $D=AB\setminus C$ is the complement of $C$ in $AB$, and $R$ is a register holding a footprint of the input. The TI compares the amount of input information that can be retrieved from $C$ and $D$ separately or jointly: if $I_3<0$, part of the information is stored in the correlations between $C$ and $D$ and it cannot be accessed locally. 

We previously showed in~\cite{lomonaco2023quantum} that the tripartite information, as discussed in~\cite{hosur2016chaos}, is a problematic quantifier of QIS in some scenarios. In particular, it is not sensible to how different choices of encoding strategies affect our ability to retrieve information from local measurements. Moreover, the tripartite information contains contributions arising from quantum discord, that may cause a significant overestimation of the actual decoding capability.
We showed in~\cite{lomonaco2023quantum} that these issues can be solved by replacing the quantum mutual informations in~\cref{eq:TI} with their accessible counterparts, and suitably performing a maximization over the choices of input encodings. This procedure gives rise to a new quantifier, dubbed \emph{accessible tripartite information} (ATI), that has the advantage of being sensible to all possible encoding-decoding strategies, and at the same time contains no discord contributions.
Computing the ATI thus requires an optimization with respect to both encoding and decoding strategies.
In this paper we introduce a new quantifier based on entropic quantities, that retains all the virtues of the ATI, maintaining in particular a clear operative interpretation, that at the same time does not require to optimize over encoding strategies, and can be computed efficiently using convex programming~\cite{skrzypczyk2023semidefinite}.


\section{Operative definition of scrambling}
\label{sec:our_definition_of_QIS}

In order to define a QIS quantifier with an immediate operative interpretation, we need to have clear in mind a realistic setup. Since QIS concerns the spreading and transmission of information, it is natural to use the typical language of quantum communication.

\subsection{QIS in the state disctimination framework}

\begin{figure}[tb]
    \centering
    \includegraphics[scale=0.35]{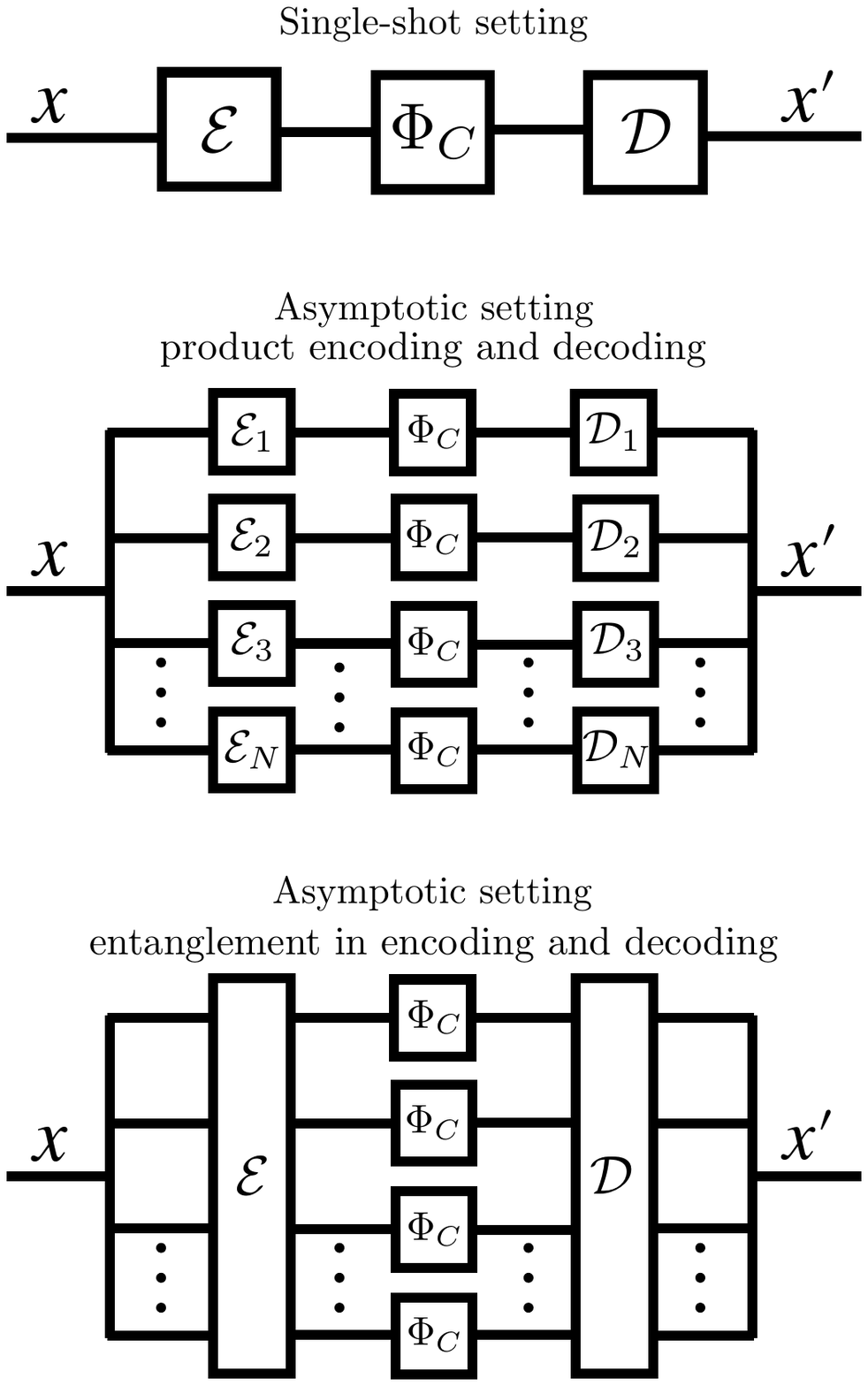}
    \caption{A generic protocol for classical communication includes three elements: an encoding step ($\mathcal{E}$) where a letter $x$ is encoded in a state $\rho_x$. The action of the channel $\Phi_{C}$ and the decoding step ($\mathcal{D}$), the measurement. Top: the single-shot setting where the channel is used a single time. Middle: the asymptotic setting, where the channel can be used $N$ times but no entanglement is exploited either in the encoding or decoding step. Bottom: the asymptotic setting with entanglement resources allowed both in input and output.
    }
    \label{fig:single-shot}
\end{figure}

Let us consider a discrimination scenario where Alice sends one of a set of quantum states through some unitary evolution, and Charlie wants to figure out which state was sent by Alice, but can only measure a fixed subset of the output state.
Formally, Alice extracts a letter $x$ with probability $p_x$ from an alphabet of $\mathcal{X}$ letters, prepares the state $\rho_x\in\mathcal{D}_A$ and keeps track of the letter in a classical register $X$.
Alice then sends $\rho_x$ through an isometric evolution $V:\mathcal{H}_A\rightarrow \mathcal{H}_{S}$~\footnote{Note that using an isometry rather than a unitary transformation here does not meaningfully change the perspective typically used in the context of QIS. Rather, it is simply a more convenient and direct method to consider how a larger unitary operation acts on the subset of inputs we are considering. Formally, we can always write
\begin{equation}
V\ket{\psi}_A\,=\,U_{S}\ket{\psi}_A\otimes\ket{0}_{S/A}
\end{equation}
for an appropriate choice of unitary $U$ and state $\ket{0}_{S \setminus A}$, with $S \setminus  A$ the complement of $A$ in $S$.},
embedding the input state into a larger system $S$.
Charlie is now tasked to retrieve $x$ by performing a generalized measurement (POVM) on some subset $C_k\subset S$ of size $k$ of the output system.
We say that $k$-scrambling occurred if there is no choice of subsystem $C_k$ and POVM that allows Charlie to recover $x$ with high probability.
It is evident that Charlie's ability to guess Alice's state is intimately related to her ability to send information through the channel $\Phi_{C_k}:\mathcal D_A\rightarrow \mathcal D_{C_k}$ defined as
\begin{equation}
\label{eq:Phi_C}
\Phi_{C_k}(\rho_A)=\Tr_{S\setminus {C_k}}\left(V\rho_A V^\dagger\right)\,.
\end{equation}
In~\cref{subsec:HminAcc}, we will show that it is possible to quantify QIS as the \textit{accessible min-information}, defined as~\cite{skrzypczyk2019robustness,takagi2019general}
\begin{equation}
\label{eq:acc_min}
I^{\rm acc}_{\min}(V,k)=\max_{C_k}\max_{\mathcal{E}}\left(H_{\min}(X)-H_{\min}(X|C_k)\right)\,,
\end{equation}
where $H_{\rm min}$ is the min-entropy~\cite{konig2009operational}, whose definition is reported in~\cref{app:entropic}.
This quantity can be understood as a generalized channel capacity of $\Phi_{C_k}$ in the single-shot regime, and quantifies the maximal amount of information that can transmitted via $\Phi$ using an optimal encoding-decoding strategy.
By ``single-shot regime'' we mean here that we do not allow for encoding and decoding strategies involving sending multiple copies of the state through the channels, which is the kind of strategies standard entropic quantities consider; rather, we consider scenarios where a single copy of input state is sent through the channel once, and the decoder can only use the single state at the output to recover $x$.
Furthermore, the reliance of $I_{\min}^{\rm acc}$ on only accessible quantities makes it free of any quantum discord contribution, and thus gives it a more direct operational interpretation.
Quantifying correlations in the single-shot regime makes for quantifiers that more closely match the intuitive understanding of QIS, as in QIS one does not generally want to consider information encoded through multiple uses of the channel $\Phi_{C_k}$ --- remembering for example the original conception of QIS to study the information recoverable from black hole radiation.
\Cref{fig:single-shot} schematically reviews the differences between single-shot and asymptotic regimes. In asymptotic regimes, multiple uses of the quantum channel are allowed. When the number of uses tends to infinity, the rate of transmitted classical information can be explicitly computed via the mutual information.
This rate is given by the accessible mutual information of the quantum channel if entanglement cannot be used, \cref{fig:single-shot}(middle), and with the quantum mutual information if entanglement resources are allowed \cref{fig:single-shot}(bottom) \cite{watrous2018theory,wilde2013quantum}.

As will be shown in~\cref{subsec:opt}, performing the optimization in~\eqref{eq:acc_min} over the ensembles $\mathcal{E}$ we obtain our main result:
\begin{equation}\small
\label{eq:robust}
I_{\min}^{\rm acc}(V,k)=\max_{\boldsymbol{\mu}_{k}}\,\log\left(\sum_x\norm{V^\dagger\left(\mathbb{I}_{S\setminus C_k}\otimes \mu_x \right)V}_\infty\right),
\end{equation}
where the maximization is over all POVMs $\bs\mu_k$ acting on any $k$-dimensional subsystem $C_k$.
This simpler expression can be efficiently computed via convex programming~\cite{watrous2018theory,skrzypczyk2023semidefinite}.
Note how in~\eqref{eq:robust} we only need to optimize with respect to POVMs, and thus in particular do not need to optimize over encoding strategies. This shift of the focus from encoding ensembles to measured observables is in the same spirit of~\cite{zanardi2022quantum,andreadakis2023scrambling}.

\subsection{Quantifying QIS with generalized channel capacities}
\label{subsec:HminAcc}
Let us add more details to our discussion to understand how the generalized capacity \eqref{eq:acc_min} emerges.

\begin{figure}[tb]
    \centering
    \includegraphics[scale=0.2]{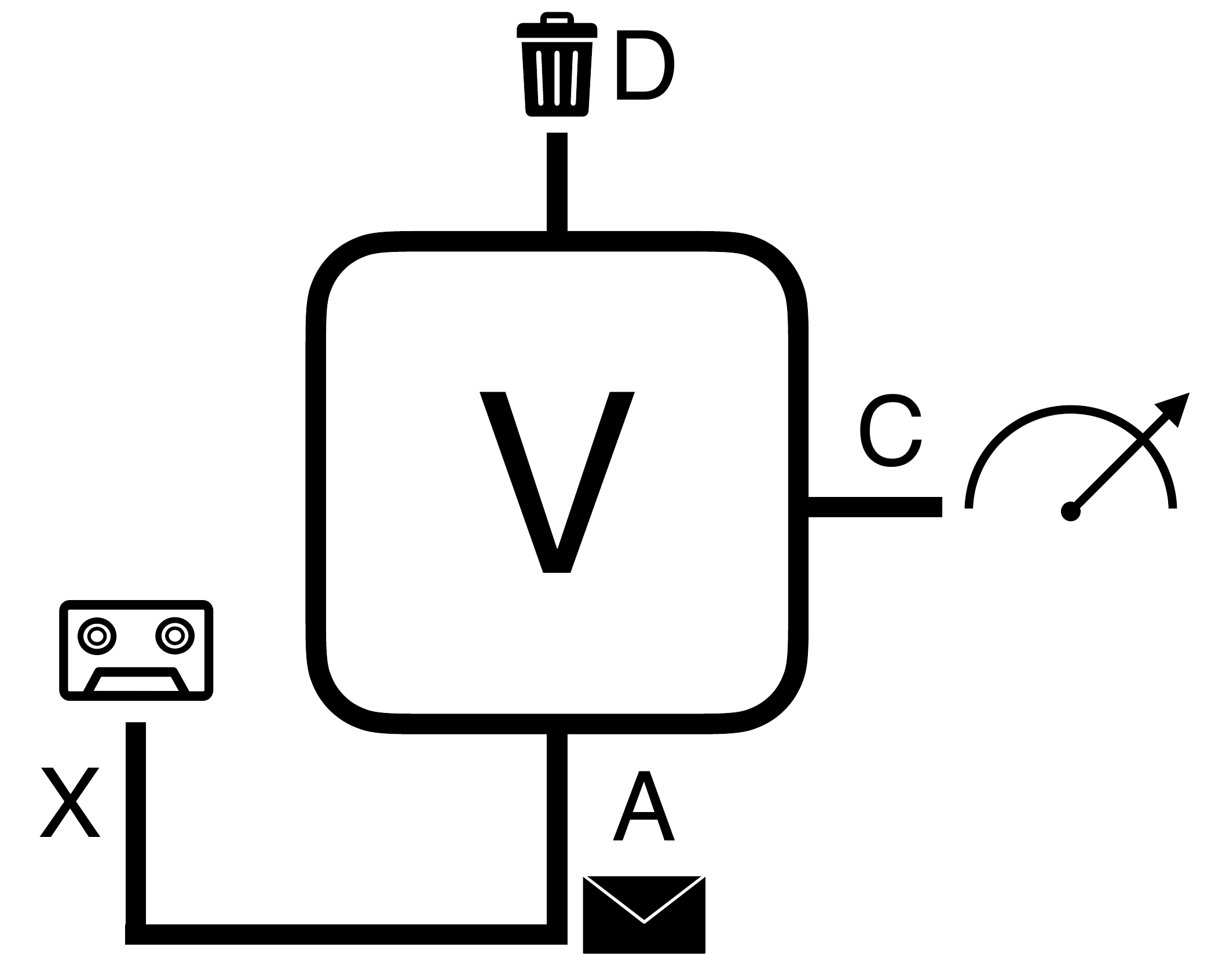}
    \caption{Schematic representation of a possible setup used to test scrambling. Alice encodes a classical message in a quantum state and keeps track of it in a classical register $X$. The state is sent through an isometric channel and the observer Charlie performs a local measurement in order to guess the sent codeword. The remaining part of the system, $D$, is inaccessible.
    }
    \label{fig:State_Discr_Setup}
\end{figure}

In the setup described at the beginning of \cref{sec:our_definition_of_QIS} and schematically represented in \cref{fig:State_Discr_Setup}, Alice draws a codeword from an ensemble $\mathcal{E}=\{(p_x,\rho_x)\}_{x=1}^\mathcal{X}$ and keeps track of the draw in classical register $X$. This amounts to preparing the classical-quantum state:
\begin{equation}
    \rho_{XA}\,=\,\sum_x p_x\,\mathbb{P}_{x}\otimes{\rho_x}
\end{equation}
where $\mathbb{P}_x=\ketbra{x}{x}$, $x=1,...,\mathcal{X}$, are projectors over an orthonormal basis for the classical register $X$. Alice embeds the message into the larger space $S$ using the isometry $V$ but only a portion $C$ can be accessed by Charlie.
The classical-quantum state encoding the message sent by Alice and the state received by Charlie can then be written as:
\begin{equation}
\label{eq:rho_XC}
\rho_{XC}=\sum_{x}p_x\mathbb{P}_x\otimes \Phi_C(\rho_x)\,.
\end{equation}
Charlie's decoding strategy is to now perform a POVM $\boldsymbol{\mu}=\{\mu_x\}_{x=1}^{\mathcal{X}}$ on $C$, and simply guess $\rho_x$ if the measurement outcome is $x$.
The probability to guess correctly is~\cite{bae2015quantum,barnett2009quantum}:
\begin{equation}
\label{eq:p_guess}
    p_{\rm guess}(\mathcal{E},\boldsymbol{\mu})\,=\,\sum_{x}p_x\Tr(\mu_x\Phi_C(\rho_x)),
\end{equation}
By optimizing this probability over all possible encoding strategies $\mathcal{E}$ and POVMs $\boldsymbol{\mu}$, one gets an estimate of Charlie's ability to retrieve the input $x$.
This procedure would however lead to a misleading result. Let us consider a scenario where a specific letter $x_0$ has a high probability of being chosen, $p_0=1-\epsilon,\,\,\epsilon\ll 1$. Alice transmits the encoded state through the channel $\Phi_C$, and Charlie attempts to guess the content of the received message. In this case, an effective strategy for Charlie would be to always guess $x_0$, resulting in a high success rate, irrespective of the details about the channel $\Phi_C$.
The triviality of such estimation strategy arises from the biased introduced by the priors $\{ p_x\}$.
One way to obtain a quantifier that is not affected by these priors in the input messages, 
is to consider instead the following ratio:
\begin{equation}
\label{eq:discriminatio_ratio}
    r_{\rm guess}(\mathcal{E},\boldsymbol{\mu})=\sum_x \frac{p_x}{p_{\max}}\Tr(\mu_x\Phi_C(\rho_x))
\end{equation}
where $p_{\max}=\max_x p_x$. This ratio approaches 1 when either all the states in the ensemble become indistinguishable after the action of the channel or when $p_{\max}\approx 1$. In contrast, it achieves its maximum value $\mathcal{X}$ if and only if two conditions are simultaneously satisfied: (1) the prior probabilities are uniform, with $p_x=\frac{1}{\mathcal{X}}$ for all $x$, and (2) all the output states are perfectly distinguishable.

We can now maximize $\eqref{eq:discriminatio_ratio}$ over the encoding-decoding strategies in order to have an estimate of the optimal Charlie's ability to retrieve the input classical information.
To this end, remember the relation between optimal discrimination probability and conditional min-entropy $H_{\rm min}(X|C)$%
~\cite{konig2009operational}:
\begin{equation}
\max_{\boldsymbol{\mu}}p_{\text{guess}}(\mathcal{E},\boldsymbol{\mu})=2^{-H_{\min}(X|C)},
\end{equation}
where $H_{\min}(X|C)$ is evaluated on the classical-quantum state $\rho_{XC}$\eqref{eq:rho_XC}. We then quantify Charlie's decoding ability as:
\begin{equation}
\label{eq:Min_Holevo}
    I^{\rm acc}_{\min}(V,k)=\max_{C_k}\max_{\mathcal{E}}\left(H_{\min}(X)-H_{\min}(X|C_k)\right)\,.
\end{equation}
Note that we need to perform an optimization over all possible subsystem $C_k$ of size $k$: this is necessary in order to reliably check whether the input information is stored in \emph{any} such subsystem.
Using the techniques discussed in~\cite{skrzypczyk2019robustness}, the accessible min-information can be shown to be maximized by ensembles with uniform prior.

\subsection{Optimization of the accessible min-information}
\label{subsec:opt}

We can simplify the handling of~\cref{eq:Min_Holevo} by explicitly working out the optimization over the ensembles $\calE$. 
We start observing that Charlie's optimal discrimination probability corresponding to an ensemble $\calE$, as given by~\cref{eq:p_guess}, can be equivalently written as 
\begin{equation}
    p_{\text{guess}}(\mathcal{E},\boldsymbol{\mu})=\sum_x p_x\Tr(\Phi^\dagger_C(\mu_x) \rho_x)\,,
\end{equation}
where $\Phi^\dagger_C:\calD_C\rightarrow \calD_A$ is the adjoint channel of $\Phi_C$, and $\{\Phi_{C}^\dagger(\mu_x)\}_{x=1,\dots,\mathcal X}$ is a POVM on $A$~\cite{watrous2018theory}.
The accessible min-information then reads
\begin{equation}
\label{eq:stepOptRetr}
    I^{\rm acc}_{\min}(V,k)=\max_{\mathcal{E},\bs{\mu}_k}\log\sum_x \frac{p_x}{p_{\max}}\Tr(\Phi^\dagger_{C_k}(\mu_x) \rho_x)\,.
\end{equation}
where the maximization is again over all POVMs $\bs\mu_k$ acting on any $k$-dimensional subsystem $C_k$.
This allows to perform the maximization over $\calE$ explicitly using the following identity, valid for any POVM $\boldsymbol{\eta}$ \cite{skrzypczyk2019robustness}:
\begin{equation}
\max_{\mathcal{E}} \sum_x\frac{p_x}{p_{\max}}\Tr\left( \eta_x \rho_x\right)=\sum_x\norm{\eta_x}_\infty\equiv \mathcal{R}(\boldsymbol{\eta})+1,
\end{equation}
where $\norm{\cdot}_\infty$ is the operator norm, and $\mathcal{R}(\boldsymbol{\eta})$ is the \emph{robustness of measurement}, a quantity shown in~\cite{skrzypczyk2019robustness} to equal the minimal amount of ``noise'' that, added to $\bs\eta$, gives a trivial measurement.
We then rewrite $I_{\rm min}^{\rm acc}$ as
\begin{equation}
\label{eq:OptRetr}
\begin{split}
    I_{\min}^{\rm acc}(V,k)=&\max_{\bs\mu_k}\,\log\left(\sum_x\norm{\Phi^\dagger_{C_k}(\mu_x)}_\infty\right)\\
    =& \max_{\bs\mu_k}\log\left(1+\mathcal{R}(\Phi^\dagger_{C_k}\left(\boldsymbol{\mu})\right)\right) \,,
\end{split}    
\end{equation}
where $\Phi^\dagger_{C_k}(\bs\mu)$ is the POVM whose elements are the operators $\Phi^\dagger_{C_k}(\mu_x)$.
Computing the accessible min-information thus reduces to an optimization over POVMs acting on $k$-dimensional subsystems.
More specifically, being the operator norm $\|\cdot\|_\infty$ a convex function, computing $I^{\rm acc}_{\min}(V,k)$ is a convex program~\cite{watrous2018theory,skrzypczyk2023semidefinite}.
Because the accessible min-information is a convex function over POVMs \cite{skrzypczyk2019robustness}, it can be shown that the optimal POVM has at most $k\leq \mathcal{X}\leq k^2$ effects~\cite{watrous2018theory}.

An isometry $V$ is a perfect $k$-scrambler if and only if the induced channels $\Phi_{C_k}$ completely erase the input information for any $k$-dimensional subsystem $C_k$ --- that is, they are replacement channels: $\Phi_{C_k}(\rho_A)=\trace(\rho_A) \sigma_k$ for some state $\sigma_k$.
This condition is equivalent to the adjoints $\Phi_{C_k}^\dagger$ sending any POVM $\bs\mu_k$ to a trivial measurement, meaning $\Phi^\dagger_{C_k}(\mu_x)\propto\mathbb I_A$ $\forall x$, and therefore to the robustness $\mathcal R(\Phi^\dagger_{C_k}(\bs\mu))$ being zero~\cite{skrzypczyk2019robustness}.
We thus conclude that an isometry $V$ is a perfect $k$-scrambler if and only if $I_{\rm min}^{\rm acc}(V,k)=0$.

To check that this condition holds for any POVM it is sufficient to verify it for a local IC-POVM on each subsystem $C_k$.
For instance, if $C$ consists of a single qubit, it is enough to construct an isometry $V$ such that all $\Phi^\dagger_{C_k}$ map the single-qubit SIC-POVM to the trivial measurement on $A$.

\begin{figure*}[t]\centering
	\includegraphics[width=\linewidth]{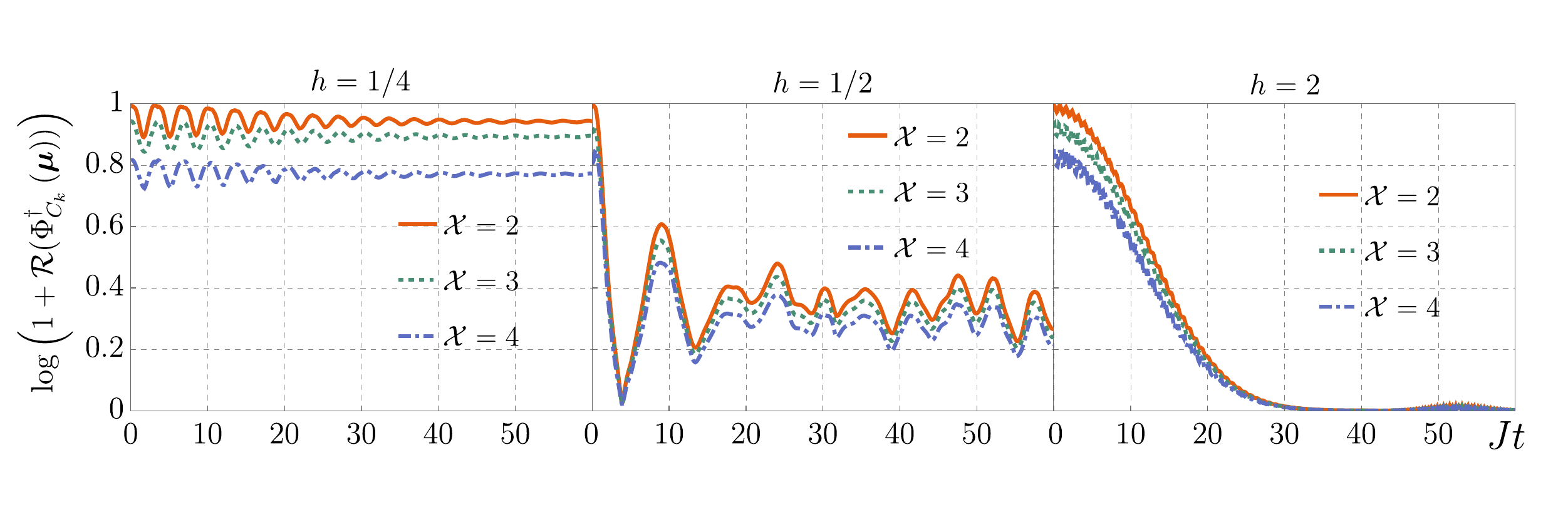}
    \caption{Time evolution in the LMG model of $\log\left(1+\mathcal{R}(\Phi^\dagger_{C_k}\left(\boldsymbol{\mu})\right)\right)$, when optimized over POVMs with a fixed number of elements $\mathcal{X}=2,3,4$. The accessible min-information at each time $t$ coincides in turn with the maximum over $\mathcal{X}$. Moving from the left to the right, the same analysis has been performed for different values of the coupling $h$. We can observe three different behaviours crossing the dynamical phase transition at $h=1/2$. Information scrambling occurs only for $h\geq 1/2$. However, only in the supercritical phase the accessible min-information remains  close to zero at large times, while at the critical point we can observe a chaotic behaviour. In all the phases, the optimal POVM maximizing the robustness has two elements at each time.
    }
    \label{fig:LMG-h}
\end{figure*}

\section{Applications}
\subsection{Quantum error correction codes}
\label{sec:QECC}
In the context of fault-tolerant quantum computation, quantum error correction codes (QECC) can be represented by an isometry $V:\mathcal{H}_{A}\rightarrow\mathcal{H}_{S}$ from a \emph{logical} space, $A$, to a larger \emph{physical} space, $S$.
A $[[n,j,t+1]]$-code is defined as a QECC that encodes the information stored in $j$ $d$-dimensional qudits of $A$ into $n$ qudits of $S$ with a minimum distance of $t+1$ between any two codewords in $S$, equal to the minimum number of qudits that must be affected by an error to irretrievably corrupt the encoded information.
It can be shown~\cite{preskill1998lecture} that \emph{any} $t$-qudit subsystem $C$ of the physical space of a $[[n,j,t+1]]$-code induces a channel $\Phi_C$ that is a replacement channel.
Therefore, as previously discussed, its adjoint maps any POVM to a trivial measurement, and thus $I_{\rm min}^{\rm acc}(V,d^t)=0$ for any isometry $V$ that implements such a QECC.
We thus conclude that any $[[n,j,t+1]]$-code is a \emph{perfect} $t$-scrambler, which is an interesting venue of future research in the context of approximate or holographic QECC.

\subsection{Spin chains}
\label{sec:Chains} 
 


In this section, we apply \eqref{eq:robust} to study quantum information scrambling in the Lipkin-Meshkov-Glick (LMG) model, described by the Hamiltonian
\begin{equation}
    H_{\rm LMG}\,=\,-\frac{2J}{N}S_z^2-2hS_x\,,
\end{equation}
where $S_{x,y,z}=\frac{1}{2}\sum_{i=1}^N\sigma_{x,y,z}^{(i)}$ are the components of the total spin. Because $\mathbf{S}^2$ is conserved, $\left[H_{\rm LMG},\mathbf{S}^2\right]\,=\,0$, it is possible to study the dynamics in sectors of fixed total spin. We will focus on the sector of maximal spin, $S=\frac{N}{2}$, that includes the ground state of the model; furthermore, any eigenstate in this sector can be represented as a superposition of completely symmetric states under permutations known as Dicke states. A Dicke state $\ket{S,M}$ is an eigenstate of $\mathbf{S}^2$ with eigenvalue $S(S+1)$ and an eigenstate of $S_z$ with eigenvalue $M$. The two states of maximal spin projection $\ket{S,\pm S}$ can be written in the computational basis as $\ket{0}^N,\ket{1}^N$. The dimension of this sector is $N+1$, {\it i.e.} it scales linearly with the size of the chain, implying that the model can be studied numerically also for large values of $N$.

QIS in the LMG chain has been extensively studied in the past \cite{pappalardi2018scrambling,bhattacharjee2022krylov,xu2020does,kidd2021saddle,li2023improving}. In particular, it has been observed that the scrambling properties of the model strongly depend on the value of the transverse field coupling $h$ (in the following we will set $J=1$, {\it i.e.} we measure time in $tJ$ units.). Three different behaviours can be recognized, $h<1/2,\,h=1/2$ and $h=1/2$. In the latter case, a dynamical phase transition (DPT) occurs. In this section we show that the accessible min-information is able to capture the different scrambling properties of the $\rm LMG$ model. To address this goal, the logical qubits $\{\ket{0},\ket{1} \}$ are redundantly encoded into the Dicke states $\ket{N/2,\pm N/2}$ of an LMG chain
\begin{equation}
    V_{\rm LMG}\ket{i}=e^{-itH_{\rm LMG}}\ket{i}^N\,,\quad i=0,1\,.
\end{equation}
In order to compute the accessible min-information, we can exploit the following formula, that allows to decompose the Dicke states of a chain of size $N$ in terms of Dicke states of two smaller partitions of length $\{L,N-L\} $\cite{latorre2005entanglement}:
\begin{equation}
\small
\ket{\frac{N}{2},M}\,=\,\sum_{l=0}^Lp_{M,l}\ket{\frac{L}{2},l{-}\frac{L}{2}}\otimes\ket{\frac{N{-}L}{2},M{-}l{+}\frac{L}{2}}
\end{equation}
where the coefficients $p_{M,l}$ are given by:
\begin{equation}
    p^2_{M,l}\,=\,\frac{\binom{L}{l}\binom{N-L}{M+N/2-l}}{\binom{N}{M+N/2}}\,.
\end{equation}
In the case at hand, we can fix $L=1$. In figure \ref{fig:LMG-h}, we report the the temporal evolution of $\log\left(1+\mathcal{R}(\Phi^\dagger_{C_k}\left(\boldsymbol{\mu})\right)\right)$ when optimized over POVMs with a fixed number of elements $\mathcal{X}=2,3,4$, for three distinct values of the transverse field coupling $h$. The accessible min-information is then provided by the maximal value among the different $\mathcal{X}$. It is interesting to observe that the optimal POVM has always two elements, independently on the time and coupling.

\begin{figure}[tbh]
    \centering
    \includegraphics[scale=0.6]{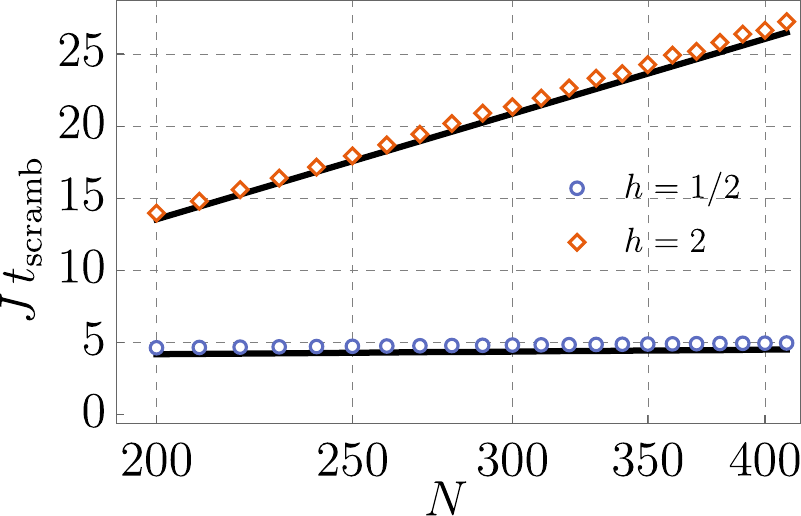}
    \caption{The behavior of the scrambling time with the size of the chain for $h=1/2$ (DPT) and $h=2$. In both cases, the scrambling time grows logarithmically with $N$. However, it is evident that at the DPT the scrambling time is attained at shorter times. Moreover, the slope in the $h>1/2$ is about $10^2$ times bigger than at the DPT.
    }
    \label{fig:scrambling_time}
\end{figure}

Within the subcritical regime, $h{<}1/2$, $I_{\min}^{\rm acc}$ always remains close to one, showing no signs of information scrambling. In contrast, within the supercritical regime ($h{>}1/2$), we observe that the accessible min-information, after a specific time referred to as the {\itshape scrambling time}, is approximately zero. At the DPT $h{=}1/2$, $I_{\min}^{\rm acc}$ experiences a rapid fall, coming close to its zero. However, after the scrambling time, it subsequently rises to higher values and enters a phase of chaotic oscillations. It is also possible to investigate how the scrambling time $t_{\rm scramb}$ depends on the length of the chain (figure \ref{fig:scrambling_time}). Either in the critical and supercritical case, the growth of $t_{\rm scramb}$ is logarithmic, and thus the LMG model behaves as a {\it fast scrambler} \cite{Sekino:2008he} in the sector of spin $S=N/2$. Notably, there is a substantial difference in the slope of this growth between the two cases. Specifically, at the dynamical phase transition, the slope is approximately $10^2$ times smaller than in the regime where $h{>}1/2$.

\section{Conclusions}
We introduced a novel approach to quantify quantum information scrambling that improves on the limitations of other entropic quantifiers, and lends itself to a clear operational interpretation. Our proposal hinges on the observation that translating QIS into the language of quantum state discrimination more closely matches the idea of ``scrambling dynamics'', and therefore makes for more reliable witnesses of QIS.

More specifically, we showed that the accessible min-information is a good candidate to quantify QIS. Although computing this quantity would involve optimizing over both encoding and decoding strategies, by leveraging results established in the context of single-shot entropies~\cite{skrzypczyk2019robustness}, we can carry out the optimization over encodings explicitly, leading to a more compact and elegant formulation that only requires an optimization of the \textit{robustness of measurement}~\cite{skrzypczyk2019robustness} over local POVMs. The explicit application of our proposal to spin chains seems to suggest that this optimization can be further constrained. This approach finds a natural application in the study of quantum error correction codes, which we showed to be instances of perfect scramblers.
Building solid foundations to study the information scrambling of general isometric dynamics makes it directly amenable to experimental verification with, for example, quantum optical experimental setups where information encoded into the polarization degree of freedom is distributed into a larger orbital angular momentum space --- properly understanding the scrambling properties of such dynamics would allow to better understand the types of dynamics that allow to perform state reconstruction from local measurements~\cite{suprano2021dynamical,zia2023regression,innocenti2023potential}.
More generally, our work paves the way for a more thorough understanding of the relations between approximate and holographic quantum error correction codes and quantum information scrambling.

\newpage
\acknowledgments
LI acknowledges support from MUR and AWS under project PON Ricerca e Innovazione 2014-2020, ``calcolo quantistico in dispositivi quantistici rumorosi nel regime di scala intermedia" (NISQ - Noisy, Intermediate-Scale Quantum). GLM and GMP acknowledge funding from the European Union - NextGenerationEU through the Italian Ministry of University and Research under PNRR - M4C2-I1.3 Project PE-00000019 "HEAL ITALIA" (CUP B73C22001250006 ). 
SL and GMP acknowledge support by MUR under PRIN Project No. 2017 SRN-BRK QUSHIP.
DAC acknowledges support from the Horizon Europe EIC Pathfinder project QuCoM (Grant Agreement No.101046973). DC acknowledges support from the BMBF project PhoQuant (grant no.13N16110).

\bibliography{bibliography}

\appendix
\section{(Conditional) Min-Entropy}
\label{app:entropic}
Given a bipartite state $\rho$, its {\it conditional min-entropy} can be defined as follows:
\begin{equation}
    H_{\min}(A|B)|_\rho=-\!\!\!\!\inf_{\sigma_B\in \mathcal{D}_B}\!\!\!D_{\max}(\,\rho\,\Vert\,\mathbb{I}_A\otimes \sigma_B),
\end{equation}
where $\mathcal{D}_B$ is the set of density matrices on the subsystem $B$ and:
\begin{equation}
    D_{\max}(\rho\,\Vert\,\sigma)=\inf\left\{\lambda \in \mathbb{R}\,:\rho\leq 2^\lambda\,\sigma\right\}
\end{equation}
The min-entropy can be computed through a convex programming routine as follows:
\begin{equation}
\begin{split}
    H_{\min}(A|B)|_{\rho_{AB}}=-&\min_{\sigma_B\geq0}\log\Tr{\sigma_B}\\
    &\mathbb{I}_A\otimes \sigma_B\geq \rho_{AB}
\end{split}
\end{equation}

In some particular case, the optimization can be explicitly carried out. When $\rho_{AB}$ is a product state, $\rho_{AB}=\rho_A\otimes \rho_B$, then $H_{\min}(A|B)|_{\rho_{AB}}=-\log\norm{\rho_A}_\infty$. Observe that in the product state case, the conditional min-entropy is actually independent on the details on $B$. We can thus define the (unconditional) min-entropy as:
\begin{equation}
    H_{\min}(A)|_{\rho_A}=-\log\norm{\rho_A}_\infty\,.
\end{equation}
Let us remember that $\norm{\rho_A}_\infty$ is nothing but the eigenvalue of $\rho_A$ with highest modulus.

\end{document}